 \documentclass[pmlr,twocolumn,10pt]{jmlr} 

\usepackage{natbib}
\usepackage{graphicx}
\usepackage{multirow} 
 \usepackage{rotating}
 \usepackage{longtable}

\usepackage{booktabs}
\usepackage{siunitx}

\newcommand{\equal}[1]{{\hypersetup{linkcolor=black}\thanks{#1}}}

\theorembodyfont{\upshape}
\theoremheaderfont{\scshape}
\theorempostheader{:}
\theoremsep{\newline}

\jmlrvolume{LEAVE UNSET}
\jmlryear{2023}
\jmlrsubmitted{LEAVE UNSET}
\jmlrpublished{LEAVE UNSET}
\jmlrworkshop{Machine Learning for Health (ML4H) 2023} 

\title[Deep Phenotyping NAFLD]{Deep Phenotyping of Non-Alcoholic Fatty Liver Disease Patients with Genetic Factors for Insights into the Complex Disease}

\author{
\Name{Tahmina Sultana Priya} 
\Email{tpriya@vt.edu}
\\
\addr Department of Computer Science, Virginia Tech, Virginia, USA
\AND
\Name{Fan Leng}
\Email{Leng.Fan@mayo.edu}
\\
\addr Data Analytics and Integration, Mayo Clinic, Rochester, USA
\AND
\Name{Anthony C. Luehrs} 
\Email{Luehrs.Tony@mayo.edu}
\\
\addr Division of Computational Biology, Mayo Clinic, Rochester, USA
\AND
\Name{Eric W. Klee}
\Email{Klee.Eric@mayo.edu}
\\
\addr Division of Computational Biology, Mayo Clinic, Rochester, USA
\AND
\Name{Alina M. Allen} 
\Email{Allen.Alina@mayo.edu}
\\
\addr Division of Gastroenterology and Hepatology, Mayo Clinic, Rochester, USA
\AND
\Name{Konstantinos N. Lazaridis} 
\Email{lazaridis.konstantinos@mayo.edu}
\\
\addr Division of Gastroenterology and Hepatology, Mayo Clinic, Rochester, USA
\AND
\Name{Danfeng (Daphne) Yao} \equal{co-corresponding authors} \Email{danfeng@vt.edu}
\\
\addr Department of Computer Science, Virginia Tech, Virginia, USA
\AND
\Name{Shulan Tian} \footnotemark[1] 
\Email{Tian.Shulan@mayo.edu}
\\
\addr Division of Computational Biology, Mayo Clinic, Rochester, USA
}

\begin{document}
\maketitle

\begin{abstract}
Non-alcoholic fatty liver disease (NAFLD) is a prevalent chronic liver disorder characterized by the excessive accumulation of fat in the liver in individuals who do not consume significant amounts of alcohol, including risk factors like obesity, insulin resistance, type 2 diabetes, etc. We aim to identify subgroups of NAFLD patients based on demographic, clinical, and genetic characteristics for precision medicine. The genomic and phenotypic data (3,408 cases and 4,739 controls) for this study were gathered from participants 
in Mayo Clinic Tapestry Study (IRB\#19-000001) and their electric health records, including their demographic, clinical, and comorbidity data, and the genotype information through whole exome sequencing performed at Helix using the Exome+\textsuperscript{\textregistered} Assay according to standard procedure (\href{https://www.helix.com/}{www.helix.com}). Factors highly relevant to NAFLD were determined by the chi-square test and stepwise backward-forward regression model. Latent class analysis (LCA) was performed on NAFLD cases using significant indicator variables to identify subgroups. The optimal clustering revealed 5 latent subgroups from 2,013 NAFLD patients (mean age 60.6 years and 62.1\% women), while a polygenic risk score based on 6 single-nucleotide polymorphism (SNP) variants and disease outcomes were used to analyze the subgroups. The groups are characterized by metabolic syndrome, obesity, different comorbidities, psychoneurological factors, and genetic factors. Odds ratios were utilized to compare the risk of complex diseases, such as fibrosis, cirrhosis, and hepatocellular carcinoma (HCC), as well as liver failure between the clusters. Cluster 2 has a significantly higher complex disease outcome compared to other clusters.

\end{abstract}

\begin{keywords}
Fatty liver disease; Polygenic risk score; Precision medicine; Deep phenotyping; NAFLD comorbidities; Latent class analysis.
\end{keywords}

\section{Introduction}
\label{sec:intro}
Fatty liver disease (FLD) is a common disease caused
by excessive fat buildup in the liver. Based on the histological characteristics, FLD can be divided into
two categories: alcoholic and nonalcoholic fatty liver
disease. NAFLD affects approximately 25\% of the population \citep{younossi2016global} and is increasingly prevalent as heterogeneous, chronic, and complex liver diseases  \citep{younossi2016global, brar2019alcoholic, mato2019biomarkers}, regardless of ethnicity, age, and sex \citep{zhou2019unexpected, doycheva2017nonalcoholic, younossi2020epidemiology}. Their impact is more significant than previously believed \citep{tilg2021nafld}. The diseases can advance from a state of simple steatosis to steatohepatitis (around 20\% of individuals)\citep{vos2020global}, with or without fibrosis. If left untreated, they can further progress to cirrhosis, liver failure, and hepatocellular carcinoma \citep{benedict2017non, yeh2014pathological, vos2020global}.

A complicated and multilayered dynamic interaction of numerous factors, such as sex \citep{mauvais2020sex}, the presence of several genetic variations \citep{trepo2020update}, the coexistence of various comorbidities \citep{byrne2015nafld}, the composition of the microbiota \citep{sharpton2021current} and others, results in the clinical presentation of NAFLD. Although metabolic dysfunction, obesity, or excess weight are frequently the root causes \citep{chalasani2012diagnosis, targher2020nafld, marchesini2001nonalcoholic, arrese2021insights}, people of Chinese descent who do not have metabolic syndrome have been found to have a greater impact on the accumulation of fat in the liver when they carry the gene patatin-like phospholipase domain containing 3 (PNPLA3), which is strongly linked to NAFLD \citep{fan2017new, shen2014pnpla}. A new acronym formed metabolic (dysfunction)-associated fatty liver disease (MAFLD). Because of the disease's high heterogeneity and complexity \citep{wang2022cardiovascular}, a one-size-fits-all treatment approach is no longer suitable \citep{eslam2020mafld}. To provide effective treatment, it is crucial to tailor the approach according to the individual's unique phenotypic and genetic information.

Deep phenotyping is defined as the precise and comprehensive characterization of patients' phenotypic abnormalities, often involving the stratification of patients into disease subclasses \citep{robinson2012deep}. Electronic health records (EHR) have been used to create comorbidity networks \citep{glicksberg2016comparative}, identify disease subgroups \citep{li2015identification}, and predict disease outcomes \citep{abraham2022dense, norgeot2019assessment}. Machine learning (ML) has been widely recognized as a scalable approach for identifying patient phenotypes. Applying machine learning in the field of NAFLD is emerging \citep{sood2018study, kantartzis2023clustering, ye2022novel}. However, deep phenotyping studies that analyze complex comorbidities and genetic factors remain limited. 

In this study, we aim to:
{\em i)} use phenotype and genotype data to identify distinct subgroups associated with NAFLD, {\em ii)} explore probabilistic methods (namely, latent class analysis) to identify subcohorts, and {\em iii)} gain insights about comorbidities in each subgroup. Our analysis identified five distinct subgroups of NAFLD patients: subgroup 1 (Non-Obese Metabolic NAFLD), subgroup 2 (Elevated NAFLD with High Genetic Risk), subgroup 3 (Metabolic-Multi-Morbid NAFLD with Psychoneurological
Burden), subgroup 4 (Male Dominant Cardiorenal NAFLD), and subgroup 5 (Non-Metabolic NAFLD). Patients in subgroup 4 had significantly higher risks of cirrhosis, liver transplantation, and HCC than those in other subgroups. Subgroup 2, characterized by genetic predisposition, is significantly associated with cirrhosis, which is consistent with previous findings \citep{liu2022health} showing risk allele carriers were at higher risk of hepatic progression rather than cardiometabolic complications.

\section{Methods \& Datasets}


\noindent
\subsection{{\bf Datasets.}} 
This work was conducted on data extracted from patients' EHR, including their demographic details, clinical variables, characterized comorbidities, and genotype information from participants in the Mayo Clinic Tapestry study, with approval from the Institutional Review Board (IRB). We identified 3,408 patients with FLD (60.7\% females, 60.2 mean age (13.1 standard deviations)) and 4,739 control patients (79.6\% females, 43.7 mean age, 14.9 standard deviations) in a population genomics study (\ref{apd:Appendix}ppendix Table A1). We collected sociodemographic information on gender, age, income, education, race, and ethnicity. Anthropometric measurements include recent height, weight, and BMI. Clinical variables include C-reactive protein (CRP), high-density and low-density lipoprotein (HDL and LDL), cholesterol, triglycerides (TG), glucose fasting (GF), hemoglobin A1C (HbA1C), blood urea nitrogen (BUN), and hepatitis B core (HBC). Liver function tests include aspartate transaminase (AST), alanine transaminase (ALT), alkaline phosphatase (ALP), gamma-glutamyl transferase (GGT), and prothrombin time (PT). For related comorbidities, we extracted the patient’s diagnosis of obesity, hyperlipidemia, metabolic syndrome (MetS), cardiovascular disease (CVD), diabetes, sleep apnea, osteoarthritis (OA), depression, gastroesophageal reflux disease (GERD), migraine, hypertension, chronic kidney disease (CKD), and neurological problems by using the tenth revision of the International Statistical Classification of Diseases and Related Health Problems (ICD-10). Natural language processing was used to further refine diagnosis in patients' clinical notes. We selected six common single nucleotide polymorphism (SNP) variants identified through genome-wide association studies \citep{Miao-NAFLD-PRS-2022} to calculate the polygenic risk score (PRS). PNPLA3 rs738409, PNPLA3 rs2294918, TM6SF2 rs58542926, SAMM50 rs2143571, GATAD2A rs4808199, and NCAN rs2228603 genotypes were coded as dosages 0, 1, and 2 for noncarriers, heterozygous carriers, and homozygous carriers of the risk-increasing allele, respectively. For each participant, the PRS was calculated as the sum of these SNP dosages (range, 0–12) (\ref{apd:Appendix}ppendix Figure A1). In preparation for the LCA computation, PRS values were divided into three categories, classified as low, middle, and high PRS \citep{park2023association} where the PRS was 0 to 2, 3 to 4, and $\geq 6$ using 40\%, 50\% to 80\%, and 90\% quantile values, respectively (\ref{apd:Appendix}ppendix Table A2). Indeed, due to data unavailability, our analysis does not include lifestyle factors such as smoking history, drinking history, as well as family history. For the available data, only significant features were included and significance was determined using stepwise regression. Race, ethnicity, and education were excluded due to their low statistical significance. We also excluded features with high missing values, e.g., HBC, GF, and PT. All continuous values were categorized into binary indicators (\ref{apd:Appendix}ppendix Table A3). The normal range of the clinical variables was collected from the Mayo Clinic website.


\noindent
\subsection{{\bf Subgroup identification algorithm.}} The task of patient stratification requires an unsupervised clustering method, because of the lack of ground truth labels. In medical literature, latent class analysis (LCA) is the most commonly used method for patient stratification \citep{ANDREACCHI2021106739, dong2022distinct, byale2022high}. Compared to K-means clustering, literature has shown that LCA is more robust and consistent in subgroup identification\citep{sinha2021practitioner}. LCA provides statistical evidence for determining the most appropriate number of clusters -- a main advantage, compared to clustering approaches that require an arbitrarily chosen cluster number. So we performed LCA using the poLCA package \citep{JSSv042i10} in RStudio (4.3.0 [2023-04-21]) to identify latent classes within the NAFLD patients. LCA is a mixture model that divides a population into mutually exclusive and exhaustive latent classes based on the probability distributions of underlying variables. For the model selection process, up to 10 classes were assessed. Standard fit statistics were utilized to carefully choose the ultimate model, which encompassed: {\em i)} statistical metrics such as Akaike information criterion (AIC), Bayesian information criterion (BIC), consistent AIC (CAIC), adjusted BIC (aBIC) log-likelihood, likelihood ration, and entropy \citep{sinha2021practitioner, weller2020latent} where lower values of BIC indicated improved fit; {\em ii)} class sample sizes (all exceeding 5\%); and {\em iii)} clinical significance, ensuring interpretability and alignment with established scientific knowledge \citep{ANDREACCHI2021106739, doi:10.1080/10705511.2013.824781}. Characteristics were organized into clusters by employing the vector assignment of predicted class memberships. 

\section{Results}
\begin{table}[]
\caption{Model-fit indices for the LCA model. We chose the model with 5 classes.}
\label{table:Table1}
\resizebox{\columnwidth}{!}{%
\begin{tabular}{|l|l|l|l|l|l|l|}
\hline
\begin{tabular}[c]{@{}l@{}}No. of \\ Class\end{tabular} &
  log-likelihood &
  BIC &
  aBIC &
  cAIC &
  likelihood-ratio &
  Entropy \\ \hline
2  & -21949.82 & 44318.04 & 44143.30 & 44373.04 & 13785.18 & 0.735 \\ \hline
3  & -21734.78 & 44100.97 & 43837.28 & 44183.97 & 13355.11 & 0.664 \\ \hline
4  & -21567.85 & 43980.13 & 43627.47 & 44091.13 & 13021.25 & 0.705 \\ \hline
\textbf{5} &
  \textbf{-21448.91} &
  \textbf{43955.25} &
  \textbf{43513.64} &
  \textbf{44094.25} &
  \textbf{12783.37} &
  \textbf{0.761} \\ \hline
6  & -21362.00 & 43994.44 & 43463.87 & 44161.44 & 12609.55 & 0.76  \\ \hline
7  & -21288.66 & 44060.75 & 43441.22 & 44255.75 & 12462.86 & 0.738 \\ \hline
8  & -21218.50 & 44133.44 & 43424.96 & 44356.44 & 12322.54 & 0.679 \\ \hline
9  & -21162.08 & 44233.61 & 43436.16 & 44484.61 & 12209.70 & 0.7   \\ \hline
10 & -21115.00 & 44352.47 & 43466.07 & 44631.47 & 12115.55 & 0.594 \\ \hline
\end{tabular}%
}
\end{table}
\subsection{{\bf Study Cohort.}}
A total of 2,013 (59.0\%) patients were eligible for analysis from 3,408 cases after the exclusion criteria (Non-NAFLD=238) and removing missing values (N=1,157) (\ref{apd:Appendix}ppendix Figure A2). The mean age of cases was 60.6 years (18-100 years (13.2 standard deviation)). In NAFLD cases, the majority of participants were female (62.1\%), compared to non-NAFLD cases (42.4\%). NAFLD patients had greater metabolic syndrome, obesity, and comorbidities like hyperlipidemia, hypertension, as well as neurological disorders. 
\\
\subsection{{\bf NAFLD Subgroups Using LCA.}}
In order to determine the optimal number of clusters and the best fit for the model, multiple solutions were tested and statistically evaluated. 
The LCA included the top 16 indicators (p$<$.001) using stepwise regression \citep{dong2022distinct}: Gender,  hyperlipidemia, MetS, sleep apnea, hypertension, CKD, A1C, ALT, AST, ALP, BMI, BUN, HDL, LDL, TG, and OA were selected for analysis (\ref{apd:Appendix}ppendix Table A4) and the best solution was achieved with 5 clusters. \autoref{table:Table1} summarizes the model fit indices. The choice of this cluster size was mainly based on the\textit{ i) }BIC value, which indicates the model’s fit, and \textit{ii)} entropy, which signifies the degree of class separation. A low BIC value and high entropy are more desirable. We found that a class size of 5 gives the best performance in terms of BIC (43955.25) and entropy (0.761). \autoref{table:Table2}  summarizes the cluster characteristics. The initial prior probabilities were randomly assigned for each cluster by the software. Upon implementing LCA, we calculated the mean prior probabilities for the latent classes, resulting in values of 0.192, 0.076, 0.245, 0.249, and 0.238, respectively, where lower probabilities suggest sparser class assignments. Cluster 2 appears more sparse than the other classes.  

{\bf Non-Obese Metabolic NAFLD}. This cluster comprised 18\% (n = 361) of the NAFLD cases and was more likely to be older (65.2 mean age) and female. These patients had high metabolic syndrome (99.7\%), hyperlipidemia (87.5\%), and OA (64.0\%) compared to other clusters. However, these patients were less likely to be obese (37.9\%) or to have diabetes, hypertension, or abnormal lab values, though they had an increased risk factor for migraine and GERD.
\begin{table*}[ht]
\caption{\text{Baseline characteristics according to LCA-derived classes}}
\label{table:Table2}
\resizebox{\linewidth}{!}{%
\begin{tabular}{|l|l|l|l|l|l|l|l|}
\hline
 &
  \textbf{} &
  \textbf{\begin{tabular}[c]{@{}l@{}}Non Obese \\ Metabolic \\ NAFLD\\ (N=361)\end{tabular}} &
  \textbf{\begin{tabular}[c]{@{}l@{}}Elevated NAFLD \\ with High \\ Genetic Risk\\ (N=154)\end{tabular}} &
  \textbf{\begin{tabular}[c]{@{}l@{}}Metabolic-Multi-Morbid \\ NAFLD with \\ Psyconeurological \\ Burden\\ (N=507)\end{tabular}} &
  \textbf{\begin{tabular}[c]{@{}l@{}}Male Dominant \\ Cardiorenal\\ NAFLD\\ (N=505)\end{tabular}} &
  \textbf{\begin{tabular}[c]{@{}l@{}}Non Metabolic \\ NAFLD\\ (N=486)\end{tabular}} &
  \textbf{\begin{tabular}[c]{@{}l@{}}Total \\ (N=2013)\end{tabular}} \\ \hline
  \textbf{Age (Years)} &
  Mean (SD) &
  \textbf{65.2 (11.2)} &
  59.4 (13.1) &
  60.1 (12.3) &
  64.6 (11.9) &
  \textbf{53.8 (13.9)} &
  60.6 (13.2) \\ \hline
  \textbf{Gender} &
  Female &
  272 (75.3\%) &
  \textbf{83 (53.9\%)} &
  \textbf{507 (100\%)} &
  \textbf{0 (0\%)} &
  389 (80.0\%) &
  1251 (62.1\%) \\ \hline
  \multirow{3}{*}{\textbf{BMI}} &
  Obesity &
  \textbf{137 (37.9\%)} &
  100 (64.9\%) &
  427 (84.3\%) &
  344 (68.1\%) &
  295 (60.7\%) &
  1303 (64.7\%) \\ \cline{2-8} 
   &
  Overweight &
  143 (39.6\%) &
  38 (24.7\%) &
  57 (11.2\%) &
  141 (27.9\%) &
  136 (28.0\%) &
  515 (25.6\%) \\ \cline{2-8} 
 &
  Normal &
  81 (22.4\%) &
  16 (10.4\%) &
  22 (4.3\%) &
  19 (3.8\%) &
  53 (10.9\%) &
  53 (10.9\%) \\ \hline
\textbf{Hyperlipidemia} &
  Yes &
  \textbf{316 (87.5\%)} &
  74 (48.1\%) &
  355 (70.0\%) &
    396 (78.4\%) &
  \textbf{0 (0\%)} &
  1141 (56.7\%) \\ \hline
\textbf{Diabetes} &
  Type 2 &
  119 (33.0\%) &
  67 (43.5\%) &
  \textbf{300 (59.2\%)} &
  242 (47.9\%) &
  \textbf{57 (11.7\%)} &
  785 (39.0\%) \\ \hline
\textbf{MetS} &
  Yes &
  \textbf{360 (99.7\%)} &
  117 (76.0\%) &
  499 (98.4\%) &
  496 (98.2\%) &
  \textbf{148 (30.5\%)} &
  1620 (80.5\%) \\ \hline
\textbf{CVD} &
  Yes &
  155 (42.9\%) &
  47 (30.5\%) &
  215 (42.4\%) &
  \textbf{249 (49.3\%)} &
  \textbf{95 (19.5\%)} &
  761 (37.8\%) \\ \hline
\textbf{Hypertension} &
  Yes &
  213 (59.0\%) &
  91 (59.1\%) &
  \textbf{344 (67.9\%)} &
  \textbf{380 (75.2\%)} &
  \textbf{159 (32.7\%)} &
  1187 (59.0\%) \\ \hline
\textbf{Depression} &
  Yes &
  116 (32.1\%) &
  47 (30.5\%) &
  \textbf{231 (45.6\%)} &
  143 (28.3\%) &
  170 (35.0\%) &
  707 (35.1\%) \\ \hline
\textbf{Migraine} &
  Yes &
  73 (20.2\%) &
  24 (15.6\%) &
  \textbf{117 (23.1\%)} &
  49 (9.7\%) &
  113 (23.3\%) &
  376 (18.7\%) \\ \hline
\textbf{Sleep Apnea} &
  Yes &
  130 (36.0\%) &
  58 (37.7\%) &
  \textbf{293 (57.8\%)} &
  291 (57.6\%) &
  100 (20.6\%) &
  872 (43.3\%) \\ \hline
\textbf{OA} &
  Yes &
  \textbf{231 (64.0\%)} &
  60 (39.0\%) &
  \textbf{302 (59.6\%)} &
  231 (45.7\%) &
  160 (32.9\%) &
  984 (48.9\%) \\ \hline
\textbf{Gerd} &
  Yes &
  174 (48.2\%) &
  55 (35.7\%) &
  \textbf{261 (51.5\%)} &
  184 (36.4\%) &
  144 (29.6\%) &
  818 (40.6\%) \\ \hline
\textbf{CKD} &
  Yes &
  90 (24.9\%) &
  38 (24.7\%) &
  182 (35.9\%) &
  \textbf{227 (45.0\%)} &
  \textbf{69 (14.2\%)} &
  606 (30.1\%) \\ \hline
\textbf{ALT} &
  High &
  6 (1.7\%) &
  \textbf{131 (85.1\%)} &
  34 (6.7\%) &
  50 (9.9\%) &
  44 (9.1\%) &
  265 (13.2\%) \\ \hline
\textbf{AST} &
  High &
  15 (4.2\%) &
  \textbf{152 (98.7\%)} &
  9 (1.8\%) &
  3 (0.6\%) &
  11 (2.3\%) &
  190 (9.4\%) \\ \hline
\textbf{ALP} &
  High &
  18 (5.0\%) &
  \textbf{44 (28.6\%)} &
  59 (11.6\%) &
  28 (5.5\%) &
  26 (5.3\%) &
  175 (8.7\%) \\ \hline
\textbf{BUN} &
  High &
  22 (6.1\%) &
  \textbf{17 (11.0\%)} &
  55 (10.8\%) &
  \textbf{90 (17.8\%)} &
  8 (1.6\%) &
  192 (9.5\%) \\ \hline
\textbf{LDL} &
  High &
  64 (17.7\%) &
  25 (16.2\%) &
  110 (21.7\%) &
  29 (5.7\%) &
  \textbf{109 (22.4\%)} &
  337 (16.7\%) \\ \hline
\textbf{HDL} &
  Poor &
  25 (6.9\%) &
  \textbf{93 (60.4\%)} &
  322 (63.5\%) &
  \textbf{419 (83.0\%)} &
  194 (39.9\%) &
  1053 (52.3\%) \\ \hline
\textbf{TG} &
  High &
  38 (10.5\%) &
  59 (38.3\%) &
  \textbf{331 (65.3\%)} &
  236 (46.7\%) &
  121 (24.9\%) &
  785 (39.0\%) \\ \hline
\textbf{PRS} &
  High &
  49 (13.6\%) &
  \textbf{35 (22.7\%)} &
  58 (11.4\%) &
  62 (12.3\%) &
  \textbf{75 (15.4\%)} &
  279 (13.9\%) \\ \hline
\end{tabular}
}
\end{table*}

{\bf Elevated NAFLD with High Genetic Risk}. This cluster was the smallest by proportion 8\% (n = 154) of NAFLD cases and was characterized by labs suggesting liver inflammation and dysfunction, such as elevated ALT (85.1\%), AST (98.7\%) and ALP (28.6\%). These patients had fewer comorbidities than other patients, but they exhibited a higher PRS value (22.7\%) compared to other subgroups.

{\bf Metabolic-Multi-Morbid NAFLD with Psychoneurological Burden}. This cluster comprised the largest number of patients, 25.2\% (n = 507), and all were female (100\%). These patients were defined by increased scores for psychiatric diagnoses: depression (45.6\%), or any other mental health disorder, as well as migraine (23.1\%) and sleep apnea (57.8\%). We also observe that this subgroup has a higher level of comorbidities, including obesity (84.3\%), diabetes (59.2\%), BMI($>=$40, 31.6\%), MetS (98.4\%), GERD (51.5\%), OA (59.6\%) and an increased level of TG (65.3\%).

{\bf Male Dominant Cardiorenal NAFLD}. This cluster comprised the second-largest number of patients 25\% (n = 505) with NAFLD, and all were male (100\%). CVD (49.3\%), CKD (45.0\%), hypertension (75.2\%), and hyperlipidemia (78.4\%) were more common among subgroup 4 patients, while depression and migraine were less common.

{\bf Non-Metabolic NAFLD}. Patients in subgroup 5 tended to be younger (53.8 mean age), had fewer comorbidities than others, and were unlikely to have abnormal lab values. However, they had a slightly increased PRS value (15.4\%). Subgroup 5 patients were relatively healthy compared to the other groups.

\noindent
\subsection{{\bf Insights into Complex Diseases.}} We further computed the odds ratios (\autoref{table:Table3}). With subgroup 1 as the reference, subgroup 4 was strongly associated with the highest risks for HCC (OR 11.78; 95\% CI 1.56–89.21), liver transplant (OR 7; 95\% CI 2.11–23.22), and increased risk of cirrhosis (OR
1.77; 95\% CI 1.01–3.13) among all subgroups, while subgroup 3 was associated with higher risk for liver transplantation (OR 4.14; 95\% CI 1.2–14.23) and increased risk of steatohepatitis (OR 1.33; 95\% CI 0.94-1.88). Subgroup 2 was also
strongly associated with the highest risk of cirrhosis (OR 3.87; 95\% CI 2.05–7.3),
steatohepatitis, and fibrosis, along with an elevated risk of HCC and liver transplants.
Subgroup 5 exhibited a less significant impact on complex diseases, aligning with prior research indicating that this subgroup generally demonstrates better health conditions than the others.
\noindent
\begin{table}[h!]
\caption{\text{Insights into complex diseases}}
\label{table:Table3}
\resizebox{\columnwidth}{!}{%
\begin{tabular}{|l|l|l|l|l|l|}
\hline
\textbf{Classes} &
  \textbf{Fibrosis} &
  \textbf{Cirrhosis} &
  \textbf{\begin{tabular}[c]{@{}l@{}}Hepatocellular\\ Caricinoma\end{tabular}} &
  \textbf{\begin{tabular}[c]{@{}l@{}}Liver \\ Transplant\end{tabular}} &
  \textbf{Steatohepatitis} \\ \hline
\textbf{1} &
  {\color[HTML]{CB0000} \textbf{ref}} &
  \textbf{ref} &
  {\color[HTML]{9A0000} \textbf{ref}} &
  {\color[HTML]{CB0000} \textbf{ref}} &
  {\color[HTML]{CB0000} \textbf{ref}} \\ \hline
\textbf{2} &
  \textbf{\begin{tabular}[c]{@{}l@{}}1.99 \\ (0.6,6.61)\end{tabular}} &
  {\color[HTML]{CB0000} \textbf{\begin{tabular}[c]{@{}l@{}}3.87 \\ (2.05,7.3)\end{tabular}}} &
  \textbf{\begin{tabular}[c]{@{}l@{}}4.74 \\ (0.43,52.62)\end{tabular}} &
  \textbf{\begin{tabular}[c]{@{}l@{}}3.18 \\ (0.7,14.39)\end{tabular}} &
  {\color[HTML]{CB0000} \textbf{\begin{tabular}[c]{@{}l@{}}1.67 \\ (1.06,2.63)\end{tabular}}} \\ \hline
\textbf{3} &
  \textbf{\begin{tabular}[c]{@{}l@{}}1.43 \\ (0.53,3.86)\end{tabular}} &
  \textbf{\begin{tabular}[c]{@{}l@{}}1.46 \\ (0.81,2.61)\end{tabular}} &
  \textbf{\begin{tabular}[c]{@{}l@{}}3.59 \\ (0.42,30.82)\end{tabular}} &
  {\color[HTML]{CB0000} \textbf{\begin{tabular}[c]{@{}l@{}}4.14 \\ (1.2,14.23)\end{tabular}}} &
  \textbf{\begin{tabular}[c]{@{}l@{}}1.33 \\ (0.94,1.88)\end{tabular}} \\ \hline
\textbf{4} &
  \textbf{\begin{tabular}[c]{@{}l@{}}1.69 \\ (0.64,4.43)\end{tabular}} &
  {\color[HTML]{CB0000} \textbf{\begin{tabular}[c]{@{}l@{}}1.77 \\ (1.01,3.13)\end{tabular}}} &
  {\color[HTML]{CB0000} \textbf{\begin{tabular}[c]{@{}l@{}}11.78 \\ (1.56,89.21)\end{tabular}}} &
  {\color[HTML]{CB0000} \textbf{\begin{tabular}[c]{@{}l@{}}7 \\ (2.11,23.22)\end{tabular}}} &
  \textbf{\begin{tabular}[c]{@{}l@{}}1.11 \\ (0.78,1.58\end{tabular}} \\ \hline
\textbf{5} &
  \textbf{\begin{tabular}[c]{@{}l@{}}0.99 \\ (0.34,2.88)\end{tabular}} &
  \textbf{\begin{tabular}[c]{@{}l@{}}1.25 \\ (0.69,2.29)\end{tabular}} &
  \textbf{\begin{tabular}[c]{@{}l@{}}4.5 \\ (0.54,37.54)\end{tabular}} &
  \textbf{\begin{tabular}[c]{@{}l@{}}3.02 \\ (0.85,10.79)\end{tabular}} &
  \textbf{\begin{tabular}[c]{@{}l@{}}0.91 \\ (0.63,1.32)\end{tabular}} \\ \hline
\end{tabular}%
}
\footnotesize{Only
odds ratios with p $<$ 0.05 are color coded.}\\
\end{table}

\noindent
\subsection{{\bf Summary, Limitations, and Future Work.}} Among the 2,013 well-characterized NAFLD patients, 5 unique latent clusters were identified in this study. These subgroups had different clinical characteristics and different outcomes. Two groups had fewer comorbidities and more positive outcomes. Another subgroup with a high PRS value shows several complex disease outcomes. Our findings are consistent with prior studies reporting gender-stratified NAFLD \citep{ballestri2017nafld}, lipid liver NAFLD \citep{carrillo2022phenotypes}, and a comparatively healthy subgroup with younger patients with less complex disease \citep{vandromme2019automated}. In addition, the subgroups reveal that NAFLD patients with high PRS values are at an increased risk of HCC \citep{thomas2022nafld}. Our study of heterogeneity among NAFLD patients benefited from the inclusion of personalized genetic data and a thorough utilization of EHR data, which may enable more precise prevention, diagnosis, and therapy planning. The limitations of our study are common to EHR-based projects, which often suffer missing values. Additionally we converted all of the continuous values to discrete values to make it computationally efficient which may include biases. To ensure AI trustworthiness and robustness in choosing the LCA model, an out-of-distribution (OOD) test is essential. The pre-processing, missing value imputation, subgroup interpretation, working with continuous values and performing OOD tests remain open to improvement. 

We showed that unsupervised clustering can be used to identify clinically relevant disease subgroups with distinct patterns of adverse outcomes. If prospectively validated, these disease subgroups could help guide patient management and screening initiatives. Ongoing work is on expanding the PRS risk analysis, performing additional comorbidity validations, new patient assignment on the computed subgroups and comparing LCA with other unsupervised clustering methods.

\smallskip
\noindent
\textbf{Data Availability:} Tapestry data were used under license, thus are not publicly available due to use restrictions.  Reasonable requests for Tapestry data from qualified researchers may be directed to, Dr. Konstantinos Lazaridis, \href{lazaridis.konstantinos@mayo.edu}{lazaridis.konstantinos@mayo.edu}. Access to de-identified data will require a legal agreement and the permission of Helix.\\
\noindent
\textbf{Acknowledgement:}  Genomic data were provided by the Tapestry study supported by the Mayo Clinic Center for Individualized Medicine.

\bibliography{jmlr-sample}
\appendix 
\onecolumn
\section{First Appendix}\label{apd:Appendix}

\setcounter{table}{0}
\renewcommand{\thetable}{A\arabic{table}}
\fontsize{7.5}{7.5}
\selectfont
\begin{longtable}[ht]{|l|l|l|l|l|l|l|}
\caption{Percentages of patients and controls with missing data
for each variable}
\label{table:TableA1}\\
\hline
  &&\begin{tabular}[c]{@{}l@{}}Control\\ (N=4937)\end{tabular} &
  \begin{tabular}[c]{@{}l@{}}NAFLD Case\\ (N=3170)\end{tabular} &
  \begin{tabular}[c]{@{}l@{}}Non NAFLD Case\\ (N=238)\end{tabular} &
  P-value &
  \begin{tabular}[c]{@{}l@{}}Total\\ (N=8345)\end{tabular} \\ \hline
\endhead
Age                             & Mean (SD)    & 43.7 (14.9)   & 60.1 (13.2)   & 61.8 (11.9)  & \textless{}0.001                  & 50.5 (16.4)   \\ \hline
\multirow{3}{*}{Gender}         & Female       & 3928 (79.6\%) & 1967 (62.1\%) & 101 (42.4\%) & \multirow{3}{*}{\textless{}0.001} & 5996 (71.9\%) \\ \cline{2-5} \cline{7-7} 
                                & Male         & 1009 (20.4\%) & 1202 (37.9\%) & 137 (57.6\%) &                                   & 2348 (28.1\%) \\ \cline{2-5} \cline{7-7} 
                                & Missing      & 0 (0\%)       & 1 (0.0\%)     & 0 (0\%)      &                                   & 1 (0.0\%)     \\ \hline
\multirow{7}{*}{BMI}            & Normal       & 1702 (34.5\%) & 354 (11.2\%)  & 81 (34.0\%)  & \multirow{7}{*}{\textless{}0.001} & 2137 (25.6\%) \\ \cline{2-5} \cline{7-7} 
                                & Obesity I    & 788 (16.0\%)  & 932 (29.4\%)  & 49 (20.6\%)  &                                   & 1769 (21.2\%) \\ \cline{2-5} \cline{7-7} 
                                & Obesity II   & 410 (8.3\%)   & 541 (17.1\%)  & 12 (5.0\%)   &                                   & 963 (11.5\%)  \\ \cline{2-5} \cline{7-7} 
                                & Obesity III  & 317 (6.4\%)   & 465 (14.7\%)  & 6 (2.5\%)    &                                   & 788 (9.4\%)   \\ \cline{2-5} \cline{7-7} 
                                & Overweight   & 1542 (31.2\%) & 863 (27.2\%)  & 87 (36.6\%)  &                                   & 2492 (29.9\%) \\ \cline{2-5} \cline{7-7} 
                                & Underweight  & 84 (1.7\%)    & 12 (0.4\%)    & 3 (1.3\%)    &                                   & 99 (1.2\%)    \\ \cline{2-5} \cline{7-7} 
                                & Missing      & 94 (1.9\%)    & 3 (0.1\%)     & 0 (0\%)      &                                   & 97 (1.2\%)    \\ \hline
\multirow{2}{*}{Obesity}        & No           & 4181 (84.7\%) & 1758 (55.5\%) & 198 (83.2\%) & \multirow{2}{*}{\textless{}0.001} & 6137 (73.5\%) \\ \cline{2-5} \cline{7-7} 
                                & Yes          & 756 (15.3\%)  & 1412 (44.5\%) & 40 (16.8\%)  &                                   & 2208 (26.5\%) \\ \hline
\multirow{2}{*}{Hyperlipidemia} & No           & 4937 (100\%)  & 1626 (51.3\%) & 162 (68.1\%) & \multirow{2}{*}{\textless{}0.001} & 6725 (80.6\%) \\ \cline{2-5} \cline{7-7} 
                                & Yes          & 0 (0\%)       & 1544 (48.7\%) & 76 (31.9\%)  &                                   & 1620 (19.4\%) \\ \hline
\multirow{3}{*}{Diabetes}       & No           & 4926 (99.8\%) & 2207 (69.6\%) & 165 (69.3\%) & \multirow{3}{*}{\textless{}0.001} & 7298 (87.5\%) \\ \cline{2-5} \cline{7-7} 
                                & Type 1       & 11 (0.2\%)    & 5 (0.2\%)     & 0 (0\%)      &                                   & 16 (0.2\%)    \\ \cline{2-5} \cline{7-7} 
                                & Type 2       & 0 (0\%)       & 958 (30.2\%)  & 73 (30.7\%)  &                                   & 1031 (12.4\%) \\ \hline
\multirow{2}{*}{MetS}           & No           & 4937 (100\%)  & 990 (31.2\%)  & 90 (37.8\%)  & \multirow{2}{*}{\textless{}0.001} & 6017 (72.1\%) \\ \cline{2-5} \cline{7-7} 
                                & Yes          & 0 (0\%)       & 2180 (68.8\%) & 148 (62.2\%) &                                   & 2328 (27.9\%) \\ \hline
\multirow{2}{*}{CVD}            & No           & 4303 (87.2\%) & 2131 (67.2\%) & 148 (62.2\%) & \multirow{2}{*}{\textless{}0.001} & 6582 (78.9\%) \\ \cline{2-5} \cline{7-7} 
                                & Yes          & 634 (12.8\%)  & 1039 (32.8\%) & 90 (37.8\%)  &                                   & 1763 (21.1\%) \\ \hline
\multirow{2}{*}{Hypertension}   & No           & 4209 (85.3\%) & 1538 (48.5\%) & 79 (33.2\%)  & \multirow{2}{*}{\textless{}0.001} & 5826 (69.8\%) \\ \cline{2-5} \cline{7-7} 
                                & Yes          & 728 (14.7\%)  & 1632 (51.5\%) & 159 (66.8\%) &                                   & 2519 (30.2\%) \\ \hline
\multirow{2}{*}{OA}             & No           & 3425 (69.4\%) & 1850 (58.4\%) & 185 (77.7\%) & \multirow{2}{*}{\textless{}0.001} & 5460 (65.4\%) \\ \cline{2-5} \cline{7-7} 
                                & Yes          & 1512 (30.6\%) & 1320 (41.6\%) & 53 (22.3\%)  &                                   & 2885 (34.6\%) \\ \hline
\multirow{2}{*}{Depression}     & No           & 3366 (68.2\%) & 2214 (69.8\%) & 189 (79.4\%) & \multirow{2}{*}{0.14}             & 5769 (69.1\%) \\ \cline{2-5} \cline{7-7} 
                                & Yes          & 1571 (31.8\%) & 956 (30.2\%)  & 49 (20.6\%)  &                                   & 2576 (30.9\%) \\ \hline
\multirow{2}{*}{GERD}           & No           & 4276 (86.6\%) & 2034 (64.2\%) & 197 (82.8\%) & \multirow{2}{*}{\textless{}0.001} & 6507 (78.0\%) \\ \cline{2-5} \cline{7-7} 
                                & Yes          & 661 (13.4\%)  & 1136 (35.8\%) & 41 (17.2\%)  &                                   & 1838 (22.0\%) \\ \hline
\multirow{2}{*}{Neurological}   & No           & 4927 (99.8\%) & 3142 (99.1\%) & 235 (98.7\%) & \multirow{2}{*}{0.0085}           & 8304 (99.5\%) \\ \cline{2-5} \cline{7-7} 
                                & Yes          & 10 (0.2\%)    & 28 (0.9\%)    & 3 (1.3\%)    &                                   & 41 (0.5\%)    \\ \hline
\multirow{2}{*}{Migraine}       & No           & 4041 (81.9\%) & 2643 (83.4\%) & 227 (95.4\%) & \multirow{2}{*}{0.0675}           & 6911 (82.8\%) \\ \cline{2-5} \cline{7-7} 
                                & Yes          & 896 (18.1\%)  & 527 (16.6\%)  & 11 (4.6\%)   &                                   & 1434 (17.2\%) \\ \hline
\multirow{2}{*}{Sleep}          & No           & 4377 (88.7\%) & 1997 (63.0\%) & 209 (87.8\%) & \multirow{2}{*}{\textless{}0.001} & 6583 (78.9\%) \\ \cline{2-5} \cline{7-7} 
                                & Yes          & 560 (11.3\%)  & 1173 (37.0\%) & 29 (12.2\%)  &                                   & 1762 (21.1\%) \\ \hline
\multirow{3}{*}{PRS}            & 0\% to 50\%  & 2393 (48.5\%) & 1284 (40.5\%) & 97 (40.8\%)  & \multirow{3}{*}{\textless{}0.001} & 3774 (45.2\%) \\ \cline{2-5} \cline{7-7} 
                                & 50\% to 90\% & 2126 (43.1\%) & 1445 (45.6\%) & 112 (47.1\%) &                                   & 3683 (44.1\%) \\ \cline{2-5} \cline{7-7} 
                                & Top 10\%     & 418 (8.5\%)   & 441 (13.9\%)  & 29 (12.2\%)  &                                   & 888 (10.6\%)  \\ \hline
\multirow{2}{*}{CKD}            & No           & 4656 (94.3\%) & 2345 (74.0\%) & 107 (45.0\%) & \multirow{2}{*}{\textless{}0.001} & 7108 (85.2\%) \\ \cline{2-5} \cline{7-7} 
                                & Yes          & 281 (5.7\%)   & 825 (26.0\%)  & 131 (55.0\%) &                                   & 1237 (14.8\%) \\ \hline
\multirow{4}{*}{A1C (\%)}       & Diabetes     & 14 (0.3\%)    & 507 (16.0\%)  & 32 (13.4\%)  & \multirow{4}{*}{\textless{}0.001} & 553 (6.6\%)   \\ \cline{2-5} \cline{7-7} 
                                & Normal       & 1616 (32.7\%) & 1191 (37.6\%) & 140 (58.8\%) &                                   & 2947 (35.3\%) \\ \cline{2-5} \cline{7-7} 
                                & Prediabetes  & 195 (3.9\%)   & 722 (22.8\%)  & 45 (18.9\%)  &                                   & 962 (11.5\%)  \\ \cline{2-5} \cline{7-7} 
                                & Missing      & 3112 (63.0\%) & 750 (23.7\%)  & 21 (8.8\%)   &                                   & 3883 (46.5\%) \\ \hline
\multirow{4}{*}{ALT (U/L)}      & High         & 100 (2.0\%)   & 407 (12.8\%)  & 31 (13.0\%)  & \multirow{4}{*}{\textless{}0.001} & 538 (6.4\%)   \\ \cline{2-5} \cline{7-7} 
                                & Low          & 6 (0.1\%)     & 3 (0.1\%)     & 0 (0\%)      &                                   & 9 (0.1\%)     \\ \cline{2-5} \cline{7-7} 
                                & Normal       & 2789 (56.5\%) & 2405 (75.9\%) & 206 (86.6\%) &                                   & 5400 (64.7\%) \\ \cline{2-5} \cline{7-7} 
                                & Missing      & 2042 (41.4\%) & 355 (11.2\%)  & 1 (0.4\%)    &                                   & 2398 (28.7\%) \\ \hline
\multirow{4}{*}{AST (U/L)}      & High         & 66 (1.3\%)    & 293 (9.2\%)   & 29 (12.2\%)  & \multirow{4}{*}{\textless{}0.001} & 388 (4.6\%)   \\ \cline{2-5} \cline{7-7} 
                                & Low          & 1 (0.0\%)     & 2 (0.1\%)     & 0 (0\%)      &                                   & 3 (0.0\%)     \\ \cline{2-5} \cline{7-7} 
                                & Normal       & 3015 (61.1\%) & 2538 (80.1\%) & 206 (86.6\%) &                                   & 5759 (69.0\%) \\ \cline{2-5} \cline{7-7} 
                                & Missing      & 1855 (37.6\%) & 337 (10.6\%)  & 3 (1.3\%)    &                                   & 2195 (26.3\%) \\ \hline
\multirow{4}{*}{ALP (U/L)}      & High         & 81 (1.6\%)    & 252 (7.9\%)   & 45 (18.9\%)  & \multirow{4}{*}{\textless{}0.001} & 377 (4.5\%)   \\ \cline{2-5} \cline{7-7} 
                                & Low          & 138 (2.8\%)   & 74 (2.3\%)    & 2 (0.8\%)    &                                   & 214 (2.6\%)   \\ \cline{2-5} \cline{7-7} 
                                & Normal       & 2660 (53.9\%) & 2511 (79.2\%) & 187 (78.6\%) &                                   & 5358 (64.2\%)    \\ \cline{2-5} \cline{7-7} 
                                & Missing      & 2058 (41.7\%) & 333 (10.5\%)  & 4 (1.7\%)    &                                   & 2395 (28.7\%) \\ \hline
\multirow{4}{*}{BUN (mg/dl)}    & High         & 63 (1.3\%)    & 252 (7.9\%)   & 77 (32.4\%)  & \multirow{4}{*}{\textless{}0.001} & 392 (4.7\%)   \\ \cline{2-5} \cline{7-7} 
                                & Low          & 68 (1.4\%)    & 21 (0.7\%)    & 3 (1.3\%)    &                                   & 92 (1.1\%)    \\ \cline{2-5} \cline{7-7} 
                                & Normal       & 3824 (77.5\%) & 2752 (86.8\%) & 155 (65.1\%) &                                   & 6731 (80.7\%) \\ \cline{2-5} \cline{7-7} 
                                & Missing      & 982 (19.9\%)  & 145 (4.6\%)   & 3 (1.3\%)    &                                   & 1130 (13.5\%) \\ \hline
\multirow{4}{*}{LDL (mg/dl)}    & Best         & 448 (9.1\%)   & 670 (21.1\%)  & 57 (23.9\%)  & \multirow{4}{*}{\textless{}0.001} & 1175 (14.1\%) \\ \cline{2-5} \cline{7-7} 
                                & High         & 585 (11.8\%)  & 496 (15.6\%)  & 36 (15.1\%)  &                                   & 1117 (13.4\%) \\ \cline{2-5} \cline{7-7} 
                                & Optimal      & 2592 (52.5\%) & 1515 (47.8\%) & 132 (55.5\%) &                                   & 4239 (50.8\%) \\ \cline{2-5} \cline{7-7} 
                                & Missing      & 1312 (26.6\%) & 489 (15.4\%)  & 13 (5.5\%)   &                                   & 1814 (21.7\%) \\ \hline
\multirow{4}{*}{HDL (mg/dl)}    & Best         & 1820 (36.9\%) & 716 (22.6\%)  & 75 (31.5\%)  & \multirow{4}{*}{\textless{}0.001} & 2611 (31.3\%) \\ \cline{2-5} \cline{7-7} 
                                & Better       & 917 (18.6\%)  & 618 (19.5\%)  & 53 (22.3\%)  &                                   & 1588 (19.0\%) \\ \cline{2-5} \cline{7-7} 
                                & Poor         & 923 (18.7\%)  & 1378 (43.5\%) & 98 (41.2\%)  &                                   & 2399 (28.7\%) \\ \cline{2-5} \cline{7-7} 
                                & Missing      & 1277 (25.9\%) & 458 (14.4\%)  & 12 (5.0\%)   &                                   & 1747 (20.9\%) \\ \hline

\end{longtable}

\begin{table}[h]
\caption{PRS cutoff value selection}
\label{table:TableA2}
\resizebox{\textwidth}{!}{%
\begin{tabular}{|l|lllll|lll|ll|}
\hline
Quantile &
  \multicolumn{1}{l|}{10\%} &
  \multicolumn{1}{l|}{20\%} &
  \multicolumn{1}{l|}{30\%} &
  \multicolumn{1}{l|}{40\%} &
  50\% &
  \multicolumn{1}{l|}{60\%} &
  \multicolumn{1}{l|}{70\%} &
  80\% &
  \multicolumn{1}{l|}{90\%} &
  100\% \\ \hline
PRS Score &
  \multicolumn{1}{l|}{1} &
  \multicolumn{1}{l|}{1} &
  \multicolumn{1}{l|}{2} &
  \multicolumn{1}{l|}{2} &
  3 &
  \multicolumn{1}{l|}{3} &
  \multicolumn{1}{l|}{4} &
  4 &
  \multicolumn{1}{l|}{6} &
  12 \\ \hline
Categorical Conversion &
  \multicolumn{4}{l|}{Low} &
  \multicolumn{4}{l|}{Medium} &
  \multicolumn{2}{l|}{High} \\ \hline
\end{tabular}%
}
\end{table}

\begin{table*}[h]
\caption{Continuous to categorical value conversion using different ranges}
\centering
\label{tab:TableA3}
\resizebox{\linewidth}{!}{
\begin{tabular}{|l|ll|l|ll|}
\hline
\multicolumn{1}{|c|}{Variable Name} &
  \multicolumn{2}{c|}{Range} &
  \multicolumn{1}{c|}{Variable Name} &
  \multicolumn{2}{c|}{Range} \\ \hline
\multirow{2}{*}{CRP (mg/L)}   & \multicolumn{1}{l|}{Low}  & \textless{}8       & \multirow{2}{*}{TG (md/dL)} & \multicolumn{1}{l|}{Desirable} & \textless{}150     \\ \cline{2-3} \cline{5-6} 
 &
  \multicolumn{1}{l|}{High} &
  \textgreater{}= 8 &
   &
  \multicolumn{1}{l|}{High} &
  \textgreater{}=150 \\ \hline
\multirow{3}{*}{HDL (md/dL)} &
  \multicolumn{1}{l|}{Poor} &
  \textless{}50 &
  \multirow{3}{*}{ALT (U/L)} &
  \multicolumn{1}{l|}{High} &
  \textgreater{}55 \\ \cline{2-3} \cline{5-6} 
 &
  \multicolumn{1}{l|}{Better} &
  50 to 59 &
   &
  \multicolumn{1}{l|}{Normal} &
  7 to 55 \\ \cline{2-3} \cline{5-6} 
 &
  \multicolumn{1}{l|}{Best} &
  \textgreater{}=60 &
   &
  \multicolumn{1}{l|}{Low} &
  \textless{}7 \\ \hline
\multirow{3}{*}{LDL (md/dL)}  & \multicolumn{1}{l|}{Best} & \textless{}70      & \multirow{3}{*}{HbA1C (\%)} & \multicolumn{1}{l|}{Diabetes}  & \textgreater{}=6.5 \\ \cline{2-3} \cline{5-6} 
 &
  \multicolumn{1}{l|}{Optimal} &
  70 to 128 &
   &
  \multicolumn{1}{l|}{Prediabetes} &
  5.7 to 6.5 \\ \cline{2-3} \cline{5-6} 
 &
  \multicolumn{1}{l|}{High} &
  \textgreater{}=129 &
   &
  \multicolumn{1}{l|}{Normal} &
  \textless{}5.7 \\ \hline
\multirow{3}{*}{BUN (md/dL)}  & \multicolumn{1}{l|}{High} & \textgreater{}24   & \multirow{3}{*}{GF (mg/dL)} & \multicolumn{1}{l|}{Diabetes}  & \textgreater{}=126 \\ \cline{2-3} \cline{5-6} 
 &
  \multicolumn{1}{l|}{Normal} &
  6 to 24 &
   &
  \multicolumn{1}{l|}{Prediabetes} &
  100 to 126 \\ \cline{2-3} \cline{5-6} 
 &
  \multicolumn{1}{l|}{Low} &
  \textless{}6 &
   &
  \multicolumn{1}{l|}{Normal} &
  \textless{}100 \\ \hline
\multirow{3}{*}{PT (seconds)} & \multicolumn{1}{l|}{High} & \textgreater{}12.5 & \multirow{3}{*}{AST (U/L)}  & \multicolumn{1}{l|}{High}      & \textgreater{}48   \\ \cline{2-3} \cline{5-6} 
 &
  \multicolumn{1}{l|}{Normal} &
  9.4 to 12.5 &
   &
  \multicolumn{1}{l|}{Normal} &
  8 to 48 \\ \cline{2-3} \cline{5-6} 
 &
  \multicolumn{1}{l|}{Low} &
  \textless{}9.4 &
   &
  \multicolumn{1}{l|}{Low} &
  \textless{}8 \\ \hline
\multirow{3}{*}{ALP (U/L)} &
  \multicolumn{1}{l|}{High} &
  \textgreater{}129 &
  \multirow{3}{*}{PRS} &
  \multicolumn{1}{l|}{High} &
  \textgreater{}=6 \\ \cline{2-3} \cline{5-6} 
 &
  \multicolumn{1}{l|}{Normal} &
  40 to 129 &
   &
  \multicolumn{1}{l|}{Middle} &
  \textgreater{}=3 \\ \cline{2-3} \cline{5-6} 
 &
  \multicolumn{1}{l|}{Low} &
  \textless{}40 &
   &
  \multicolumn{1}{l|}{Low} &
  \textless{}3 \\ \hline
\end{tabular}}
\end{table*}


\begin{table*}[h]
\caption{Variable parameters selection in cluster analysis}
\label{tab:TableA4}
\resizebox{\textwidth}{!}{%
\begin{tabular}{|l|l|l|l|l|l|}
\hline
 &
  Variable &
  Type of Variable &
  Scale of Measurement &
  \begin{tabular}[c]{@{}l@{}}Included for \\ LCA Indicator\end{tabular} &
  \begin{tabular}[c]{@{}l@{}}Included for \\ Cluster Analysis\end{tabular} \\ \hline
\multirow{4}{*}{Socio demographic}   & Age               & Continuous  & Years             &     & Yes \\ \cline{2-6} 
                                     & Gender            & Categorical & Male or Female    & Yes & Yes \\ \cline{2-6} 
 &
  Race &
  Categorical &
  \begin{tabular}[c]{@{}l@{}}American Indian/Alaskan, Native, Asian,\\ Black or African American,White,Other\end{tabular} &
   &
   \\ \cline{2-6} 
 &
  Ethnicity &
  Categorical &
  \begin{tabular}[c]{@{}l@{}}Hispanic or Not Hispanic\end{tabular} &
   &
   \\ \hline
\multirow{3}{*}{Anthropometric}      & Height            & Continuous  & kg                &     &     \\ \cline{2-6} 
                                     & Weight            & Continuous  & cm                &     &     \\ \cline{2-6} 
                                     & BMI               & Continuous  & —                 & Yes & Yes \\ \hline
\multirow{13}{*}{Clinical Variables} & CRP               & Categorical & mg/L              &     &     \\ \cline{2-6} 
                                     & HDL               & Continuous  & mg/dL             & Yes & Yes \\ \cline{2-6} 
                                     & LDL               & Continuous  & mg/dL             & Yes & Yes \\ \cline{2-6} 
                                     & TG                & Continuous  & mg/dL             & Yes & Yes \\ \cline{2-6} 
                                     & GF                & Continuous  & mg/dL             &     &     \\ \cline{2-6} 
                                     & HbA1C             & Continuous  & \%                & Yes & Yes \\ \cline{2-6} 
                                     & BUN               & Continuous  & mg/dL             & Yes & Yes \\ \cline{2-6} 
 &
  HBc &
  Categorical &
  \begin{tabular}[c]{@{}l@{}}Positive, Negative, Non Reactive\end{tabular} &
   &
   \\ \cline{2-6} 
                                     & AST               & Continuous  & U/L               & Yes & Yes \\ \cline{2-6} 
                                     & ALT               & Continuous  & U/L               & Yes & Yes \\ \cline{2-6} 
                                     & ALP               & Continuous  & U/L               & Yes & Yes \\ \cline{2-6} 
                                     & GGT               & Continuous  & U/L               &     &     \\ \cline{2-6} 
                                     & PT                & Continuous  & Seconds           &     &     \\ \hline
\multirow{13}{*}{Comorbidities} &
  Hyperlipidemia &
  Categorical &
  Yes or No &
  Yes &
  Yes \\ \cline{2-6} 
                                     & MetS              & Categorical & Yes or No         & Yes & Yes \\ \cline{2-6} 
                                     & CVD               & Categorical & Yes or No         &     & Yes \\ \cline{2-6} 
                                     & CKD               & Categorical & Yes or No         & Yes & Yes \\ \cline{2-6} 
                                     & Diabetes          & Categorical & Yes or No         &     & Yes \\ \cline{2-6} 
                                     & Sleep Apnea       & Categorical & Yes or No         & Yes & Yes \\ \cline{2-6} 
                                     & OA    & Categorical & Yes or No         & Yes & Yes \\ \cline{2-6} 
                                     & Depression        & Categorical & Yes or No         &     & Yes \\ \cline{2-6} 
                                     & Migraine          & Categorical & Yes or No         &     & Yes \\ \cline{2-6} 
 &
  \begin{tabular}[c]{@{}l@{}}Neurological\end{tabular} &
  Categorical &
  Yes or No &
   &
   \\ \cline{2-6} 
                                     & GERD              & Categorical & Yes or No         &     & Yes \\ \cline{2-6} 
                                     & Hypertension      & Categorical & Yes or No         & Yes & Yes \\ \cline{2-6} 
                                     & Obesity           &  Categorical           &  Yes or No                 &     & Yes \\ \hline
\multirow{7}{*}{Genotype Data} &
  \begin{tabular}[c]{@{}l@{}}PNPLA3 rs738409\end{tabular} &
  Continuous &
  SNP dosages 0,1,2 &
   &
   \\ \cline{2-6} 
                                     & PNPLA3 rs2294918  & Continuous  & SNP dosages 0,1,2 &     &     \\ \cline{2-6} 
                                     & TM6SF2 rs58542926 & Continuous  & SNP dosages 0,1,2 &     &     \\ \cline{2-6} 
                                     & SAMM50 rs2143571  & Continuous  & SNP dosages 0,1,2 &     &     \\ \cline{2-6} 
                                     & GATAD2A rs4808199 & Continuous  & SNP dosages 0,1,2 &     &     \\ \cline{2-6} 
                                     & NCAN rs2228603    & Continuous  & SNP dosages 0,1,2 &     &     \\ \cline{2-6} 
                                     & PRS               & Continuous  & —                 &     & Yes \\ \hline
\end{tabular}
}
\end{table*}
\renewcommand{\thefigure}
{A\arabic{figure}}

\begin{figure}[h]
  \centering
  \includegraphics[width=0.8\linewidth]{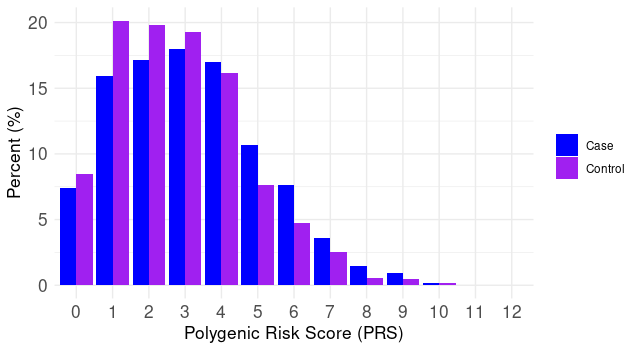}
  \caption{The distribution of polygenic risk scores of all included participants.}
  \label{app:FigureA1}
\end{figure}

\begin{figure}[h]
  \centering
  \includegraphics[width=0.65
  \linewidth]{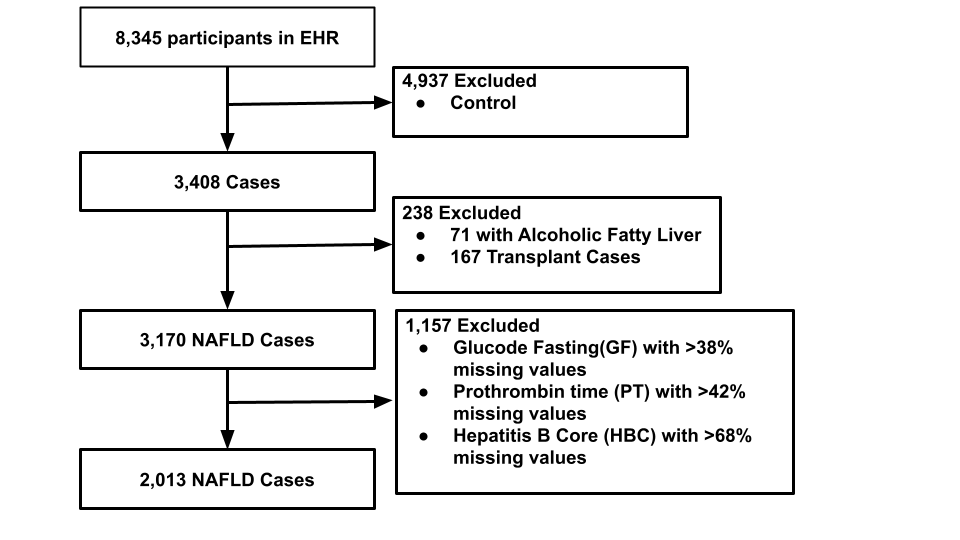}
  \caption{The study ﬂow chart.}
  \label{app:FigureA2}
\end{figure}

\end{document}